\documentclass[%
 amsmath,amssymb,
 aps,
]{revtex4-1}

\usepackage{graphicx}
\usepackage{dcolumn}
\usepackage{bm}

\begin{document}

\title{Hawking Radiation from Boundary Scalar Field}
\author{Jingbo Wang}
\email{ shuijing@mail.bnu.edu.cn}
\affiliation{Institute for Gravitation and Astrophysics, College of Physics and Electronic Engineering, Xinyang Normal University, Xinyang, 464000, P. R. China}
 \date{\today}
\begin{abstract}
Hawking radiation is one essential property of quantum black hole. It results in the information loss paradox, and give important clue to the unification of quantum mechanics and general relativity. In the previous works, the boundary scalar field on the horizon of black holes were used to give the microstates of BTZ black holes and Kerr black holes. They are account for the Bekenstein-Hawking entropy. In this paper, we show that the Hawking radiation can also be derived from those scalar fields. Actually the Hawking radiation are superposition of thermal radiation of right/left sector at different temperatures. We also discuss the impact on the information loss paradox.
\end{abstract}
\maketitle
\section{Introduction}
One important contribution of Hawking is the discovery of the Hawking radiation \cite{hawk1,hawk2}, that is, the black hole can radiate with a black body temperature $T_H$. The thermal nature of those radiation lie at the heart of the ``information loss paradox" \cite{info1,info2,info3}. After Hawking's seminal works, there are many other derivations that confirm Hawking's results. For a review, see Ref.\cite{carlip1} and references therein.

However, Hawking's analyse is semi-classical: the black hole spacetime is treated as classical background. It raises a probability that if the black hole spacetime is treated as quantum object, the thermal Hawking radiation may have modifications to contain information. Long time ago, Bekenstein \cite{bek1,bek2,bek3} has proposed that black hole play the same role in gravitations as that the atom play in quantum mechanics. This analogy suggest that the area may has a discrete spectrum,
\begin{equation}\label{1a}
  A_n=\alpha n \hbar, \quad n=1,2,3,\cdots,
\end{equation}
where $\alpha$ is dimensionless constant of order unity. Two possible values of $\alpha$ are often used: $\alpha=8\pi$ \cite{bek2,mag1}, and $\alpha=4\log k$ with $k=2,3,\cdots$ \cite{log1,bm1,log2}. The discrete area spectrum implies the discrete energy spectrum, which modify the continuous Hawking radiation to discrete line emission. Later, Bekenstein and Mukhanov \cite{bm1} suggest that the radiation of a quantum Schwarzschild black hole of mass $M$ should be at integer multiples of the fundamental frequency
\begin{equation}\label{1b}
  \omega_0=\alpha T_H=\frac{\alpha}{8\pi M}.
\end{equation}
In Ref.\cite{bm2} it suggest to observer this special frequency on the classical gravitational wave signals received by detectors.

In the previous works \cite{wangti1,wangti2}, the author claimed that the black holes can be considered as kind of topological insulators. Topological insulators \cite{ti1,ti2} are quantum many-body system. Based on this claim, we use the methods developed in topological insulators physics to study the problems in black hole physics \cite{wangbms1,wangbms2,wangplb1,wangbms4}. For example, we give the microstates for BTZ black holes and Kerr black holes \cite{wangbms4}. Those microstates can account for the Bekenstein-Hawking entropy. Due to the compactness of the scalar field, one can get the result that the area of the Kerr black hole has a discrete spectrum (\ref{1a}) with $\alpha=8\pi$. In this paper, we will show that those states can also give the Hawking radiation with some modification that similar to Bekenstein-Mukhanov's suggestion.

The paper is organized as follows. In section II, the BTZ black hole is analysed. In section III, the same method are applied to the Kerr black hole. Section IV is the conclusion.
\section{The BTZ black hole}
In this section we analyse the BTZ black hole in three dimensional spacetime. The metric of the BTZ black hole is \cite{btz1}
\begin{equation}\label{1}
    ds^2=-N^2 dv^2+2 dv dr+r^2 (d\varphi+N^\varphi dv)^2,
\end{equation}
where $N^2=-8 M+\frac{r^2}{L^2}+\frac{16 J^2}{r^2}, N^\varphi=-\frac{4 J}{r^2}$, and $(M,J)$ are mass and angular momentum respectively.

The thermodynamics quantities for BTZ black hole are
\begin{equation}\label{2}\begin{split}
  T_H=\frac{r_+^2-r_-^2}{2\pi r_+ L^2}, \quad S_{BH}=\frac{2\pi r_+}{4}, \quad \Omega_H=\frac{r_-}{r_+ L},
\end{split}\end{equation}
and satisfy the first law
\begin{equation}\label{3}
  dM=T_H dS+\Omega_H dJ.
\end{equation}
The distribution of Hawking radiation at infinity for the bosons is given by
\begin{equation}\label{4}
  <N_m (\omega)>=\frac{1}{e^{\beta(\omega-m \Omega_H)}- 1},
\end{equation}
where $\beta=\frac{1}{T}$ is the inverse temperature, $m$ is the azimuthal angular momentum number. From this distribution one can get the Planck's thermal distribution for black hole.

Now let us consider this distribution from the boundary scalar field on the horizon. The scalar field has mode expansion \cite{wangbms4}
\begin{equation}\label{4a}\begin{split}
  \phi(v',\varphi)=\phi_0+p_v v'+p_\varphi \varphi+ \sqrt{\frac{1}{m_0 A}}\sum_{n\neq 0}\sqrt{\frac{1}{2 \omega'_n}}[a_n e^{-i(\omega'_n v'-k_n \varphi)}+a^+_n e^{i(\omega'_n v'-k_n \varphi)}],\\
  \end{split}\end{equation}
where $v'=\frac{v}{\gamma}=\frac{r_+}{L}v,\omega'_n=\frac{|n|}{r_+},k_n=n$ and $A=2\pi r_+$ is the length of the circle. The scalar field $\phi(v',\varphi)$ can be considered as collectives of harmonic oscillators, and a general quantum state can be represented as $|p_v,p_\varphi;\{n_k\}>\sim (a^+_1)^{n_1}\cdots (a^+_k)^{n_k}|p_v,p_\varphi>$ where $p_v,p_\varphi$ are zero mode parts, and $\{n_k\}$ are oscillating parts. The zero mode parts give important informations about the fractional charges of the quasi-particles \cite{wangkerr4}. The oscillating parts are associated with the entropy and the radiation. The Hamiltonian and angular momentum operators have the expression
\begin{equation}\label{5}\begin{split}
  \hat{H}=\hat{H}_0+\sum_{k\neq 0} \frac{|k|}{r_+}\hat{a}_k^+ \hat{a}_k=\hat{H}_0+\sum_{k\neq 0} \varepsilon'_k \hat{n}_k,\\
   \hat{J}=\hat{J}_0 +\sum_{k\neq 0} k \hat{a}_k^+ \hat{a}_k=\hat{J}_0 +\sum_{k\neq 0} J_k \hat{n}_k, \quad k \in Z,
\end{split}\end{equation}
where $\varepsilon'_k=\frac{|k|}{r_+},J_k=k$ are energy and angular momentum for the energy level $k$, $\hat{n}_k=\hat{a}_k^+ \hat{a}_k$ is the number operator and we omit the zero-point energy. The Hawking radiation are observed at infinity, where the energy is associated with
\begin{equation}\label{7}
  \omega\sim \frac{\partial}{\partial t}=\frac{\partial}{\partial v}= \frac{\partial}{\partial v'}\frac{1}{\gamma}\sim \frac{\omega'}{\gamma}.
\end{equation}
We denote energy (temperature) that at infinity un-primed, and that on the horizon with prime $'$. They can transform into each other with scalar-factor $\gamma=\frac{L}{r_+}$.

Let us review the calculation of the entropy for BTZ black holes \cite{wangkerr4}. For the BTZ black hole with parameters $(M,J)$, the oscillating parts satisfy the constraints
\begin{equation}\label{4b}
 \frac{1}{c}\sum \frac{|k|}{r_+} n_k=M \gamma,\quad  \frac{1}{c}\sum k n_k=J, \quad n_k \in N^+, \quad c=3L/2G.
\end{equation}
Collect the positive part $k>0$ as the right sector, and negative part $k<0$ as left sector. Then they satisfy
\begin{equation}\label{4c}
  \frac{1}{c}\sum_{k>0} k n_k^R=\frac{1}{2}(M \gamma r_++J),\quad \frac{1}{c}\sum_{k<0} (-k) n_k^L=\frac{1}{2}(M\gamma r_+-J).
\end{equation}
The entropy for the right/left sector can be calculated by the Hardy-Ramanujan formula to gives
\begin{equation}\label{4d}\begin{split}
  S_R=2 \pi \sqrt{c \frac{M L+J}{12}}=\frac{\pi}{4}(r_++r_-),\\
   S_L=2 \pi \sqrt{c \frac{M L-J}{12}}=\frac{\pi}{4}(r_+-r_-).
\end{split}\end{equation}
and the total entropy is
\begin{equation}\label{4e}
  S=S_R+S_L=\frac{2\pi r_+}{4}.
\end{equation}
For right/left sectors, from the equation (\ref{4c}) we can associate them with the following energy
\begin{equation}\label{4f}
  E'_R=\frac{\gamma}{2}(M+\frac{J}{\gamma r_+}),\quad E'_L=\frac{\gamma}{2}(M-\frac{J}{\gamma r_+}).
\end{equation}
Now we can define the dimensional temperatures on the horizon for the right/left sectors by standard thermodynamics method
\begin{equation}\label{4g}
  T'_R=(\frac{\partial E'_R}{\partial S_R})_V=\frac{r_++ r_-}{2\pi L r_+}=\gamma\frac{\tilde{T}_R}{L}=\gamma T_R,\quad
  T'_L=(\frac{\partial E'_L}{\partial S_L})_V=\frac{r_+- r_-}{2\pi L r_+}=\gamma\frac{\tilde{T}_L}{L}=\gamma T_L,
\end{equation}
where $V=2 \pi r_+ c$ is the volume, $\tilde{T}_{R/L}=\frac{r_+\pm r_-}{2\pi L}$ are dimensionless temperatures. It is easy to show that those temperatures satisfy the relation
\begin{equation}\label{4h}
 \frac{2}{T_H}=\frac{1}{T_R}+\frac{1}{T_L}.
\end{equation}

The constraints (\ref{4c}) can be written as
\begin{equation}\label{5a}
  \sum_{k>0} \varepsilon'_k n_k^R=c E_R',\quad \sum_{k<0} \varepsilon'_k n_k^L=c E'_L.
\end{equation}
From statistical mechanics it is well known that the most probable distribution is the Bose-Einstein distribution
\begin{equation}\label{5b}
  <n_k^R(\varepsilon'_k)>=\frac{1}{e^{\beta'_R\varepsilon'_k}- 1},\quad <n_k^L(\varepsilon'_k)>=\frac{1}{e^{\beta'_L\varepsilon'_k}- 1}.
\end{equation}

Now we consider the radiation on the horizon. The radiations between energy levels $k,k'$ have the energy and angular momentum spectrum
\begin{equation}\label{6a}
  \omega'_n=\frac{|n|}{r_+},\quad m=n, \quad n\equiv k-k'.
\end{equation}
So for Hawking radiation at infinity, the real energy for radiations are
\begin{equation}\label{8}
  \omega_n=\frac{|n|}{r_+}\frac{1}{\gamma}=\frac{|n|}{L}.
\end{equation}
They have constant frequency spacing and have a minimal frequency
\begin{equation}\label{8a}
  \Delta \omega=\omega_0=\frac{1}{L}=\omega_{min}.
\end{equation}

Now we come to our main result: the distribution (\ref{4}) can be rewritten as follows
\begin{equation}\label{9}\begin{split}
  <N_m (\omega_n)>=\frac{1}{e^{\beta(\omega_n-m \Omega_H)}- 1}=\frac{1}{e^{\beta(\frac{|n|}{L}-n \frac{r_-}{r_+ L})}- 1}\\
 =\frac{1}{e^{\beta_R \omega_n}- 1}H(n)+\frac{1}{e^{\beta_L \omega_n}- 1}H(-n)=\frac{1}{e^{\beta'_R \omega'_n}- 1}H(n)+\frac{1}{e^{\beta'_L \omega'_n}- 1}H(-n)\\
 = <n_k^R(\omega'_n)>+ <n_k^L(\omega'_n)>,
\end{split}\end{equation}
where $H(n)$ is the Heaviside step function. That is to say, the Hawking radiation are superposition of thermal radiation (\ref{5b}) of right/left sector at different temperatures. As we said, the thermal distribution is the most probable distribution, but not the only possible distribution.

With those temperatures, the entropy and energy can be rewritten as some suggesting forms,
\begin{equation}\label{4j}
  S_{R/L}=\frac{\pi}{6} V T'_{R/L},\quad E'_{R/L}=\frac{\pi}{12} V T'^{2}_{R/L}.
\end{equation}
They are just the entropy and energy for phonon gas in one dimensional circle with length $V=2\pi r_+ c$.
\section{The Kerr black hole}
We can apply the same method to Kerr black holes. The metric of Kerr black hole can be written as \cite{kerr1}
\begin{equation}\label{20}
  ds^2=-(1-\frac{2 M r}{\rho^2})dv^2+2 dv dr-2 a \sin^2 \theta dr d\varphi-\frac{4 a M r \sin^2 \theta}{\rho^2}dv d\varphi+\rho^2 d\theta^2+\frac{\Sigma^2 \sin^2 \theta}{\rho^2}d\varphi^2,
\end{equation}
where $\rho^2=r^2+a^2 \cos^2 \theta, \Delta^2=r^2-2 M r+a^2, \Sigma^2=(r^2+a^2)\rho^2+2 a^2 M r \sin^2 \theta$.

The thermodynamics quantities for the Kerr black hole with parameters $(M,J)$ are
\begin{equation}\label{21}\begin{split}
  T_H=\frac{r_+-r_-}{8\pi r_+ M}, \quad S_{BH}=\frac{4\pi (r_+^2+a^2)}{4}, \quad \Omega_H=\frac{a}{r_+^2+a^2}.
\end{split}\end{equation}

Similar to the BTZ black hole case, the boundary can also support massless scalar field
\begin{equation}\label{30}\begin{split}
  \phi(v',\theta,\varphi)=\phi_0+p_v v'+p_\theta \ln (\cot\frac{\theta}{2})+p_\varphi\varphi+\sqrt{\frac{1}{m_0 A}}\sum_{l\neq 0}\sum_{m=-l}^{m=l}\sqrt{\frac{1}{2\omega_l}}[a_{l,m} e^{-i \omega_l v'}Y^m_l(\theta,\varphi)+a^+_{l,m} e^{i \omega_l v'}(Y^m_l)^*(\theta,\varphi)],\\
  \end{split}\end{equation}
where $\omega^2_l=\frac{l(l+1)}{r_+^2}$, $Y^m_l(\theta,\varphi)$ are spherical harmonics and $A=4\pi r_+^2$. The scalar field $\phi(v',\theta,\varphi)$ can be considered as collectives of harmonic oscillators, and a general quantum state can be represented as $|p_v,p_\varphi;\{n_{l,m}\}>$ where $p_v,p_\varphi$ are zero mode part, and $\{n_{l,m}\}$ are oscillator part.

But different from the BTZ black hole case, for Kerr black hole, the scalar field should have interaction, and the special interaction results in the particular form for the Hamiltonian and angular momentum operator \cite{wangkerr4},
\begin{equation}\label{31}\begin{split}
  \hat{H}_{full}= \hat{H}_0+\sum_m \frac{|m|}{r_+}\hat{n}_m=\hat{H}_0+\sum_{m} \varepsilon'_m \hat{n}_m, \\
  \hat{J}=\hat{J}_0+\sum_m m \hat{n}_m=\hat{J}_0 +\sum_{m} J_m \hat{n}_m, \quad m\neq 0,
  \end{split}\end{equation}
which have the same forms as those for the BTZ black hole (\ref{5}).

The microscopic states of the Kerr black hole are represented by $|0,0;\{n_m\}>$, and satisfy the constraints
\begin{equation}\label{32}
  \sum_{m\neq 0} |m| n_m=\frac{c M r_+ \gamma}{4}, \quad \sum_{m\neq 0} m n_m=\frac{c J}{2}, \quad \gamma\equiv\frac{r_+^2+a^2}{r_+^2},\quad c=12 M r_+.
\end{equation}
Separate the sequence $\{m\}$ into positive part $\{m_+\}$ with all $m_+>0$, and negative part $\{m_-\}$ with all $m_-<0$. Then the constraints become
\begin{equation}\label{33}
  \sum_{m>0} m n_m^+=\frac{\gamma}{4}(\frac{M r_+}{2}+\frac{J}{\gamma}),\quad \sum_{m<0}(-m) n_m^-=\frac{\gamma}{4}(\frac{M r_+}{2}-\frac{J}{\gamma}).
\end{equation}
For right/left sectors on the horizon, we can associate them with the following energy and entropy
\begin{equation}\label{21f}\begin{split}
  E'_R=\frac{\gamma}{4}(\frac{M}{2}+\frac{J}{\gamma r_+})=\frac{\gamma}{8}(M+a),\quad E'_L=\frac{\gamma}{8}(M-a),\\
   S_R=2 \pi \sqrt{c \frac{M^2+J}{24}}=\pi M(r_++a),\\
   S_L=2 \pi \sqrt{c \frac{M^2-J}{24}}=\pi M(r_+-a).
\end{split}\end{equation}
Now we can define the dimensional temperatures for the right/left sectors
\begin{equation}\label{21g}
  T'_R=(\frac{\partial E'_R}{\partial S_R})_V=\gamma \frac{r_++ a}{8\pi M r_+}\equiv \gamma  T_R=\frac{\tilde{T}_R}{r_+},\quad T'_L=(\frac{\partial E'_L}{\partial S_L})_V=\gamma \frac{r_+- a}{8\pi M r_+}\equiv \gamma T_L=\frac{\tilde{T}_L}{r_+},
\end{equation}
where $V=2 \pi r_+ c$ is the volume, $\tilde{T}_{L/R}=\frac{1}{4\pi}(1\pm \frac{a}{r_+})$ are dimensionless temperatures. It is easy to show that those temperatures satisfy the relation
\begin{equation}\label{21h}
 \frac{2}{T_H}=\frac{1}{T_R}+\frac{1}{T_L}.
\end{equation}

The heat capacity for right/left sectors can be calculated by
\begin{equation}\label{22}
  C_R=(\frac{\partial E'_R}{\partial T'_R})_V=S_R,\quad C_L=(\frac{\partial E'_L}{\partial T'_L})_V=S_L,
\end{equation}
which are both positive, just like ordinary matter. The crucial point is to keep the volume $V=2 \pi r_+ c=24\pi M r_+^2$ fixed. This volume is related to the geometrical volume \cite{volume1, volume2} $V'=\frac{1}{3}A_H r_H=\frac{8\pi}{3}M G r_+^2$ by $V=9 V'/G$.

Now we consider the radiation on the horizon. From the expression (\ref{31}) one can see the energy and the azimuthal angular momentum for the radiations between two levels $k,k'$ are
\begin{equation}\label{24}
  \omega'_m=\frac{|m|}{r_+},\quad m=m,\quad m\equiv k-k'.
\end{equation}

Similar to the BTZ black hole case, the energy at infinity should be
\begin{equation}\label{25}
  \omega_m=\frac{|m|}{r_+}\frac{1}{\gamma}=\frac{|m|}{2 M}.
\end{equation}
They have constant frequency spacing and a minimal frequency
\begin{equation}\label{25a}
  \Delta \omega=\omega_0=\frac{1}{2 M}=\omega_{min}.
\end{equation}

Now we can rewrite the distribution as follows
\begin{equation}\label{34}\begin{split}
  <N_m (\omega_m)>=\frac{1}{e^{\beta(\omega_m-m \Omega_H)}- 1}=\frac{1}{e^{\beta(\frac{|m|}{2 M}-m \frac{a}{r_+^2+a^2})}- 1}\\
  =\frac{1}{e^{\beta_R \omega_m}- 1}H(m)+\frac{1}{e^{\beta_L \omega_m}- 1}H(-m)=\frac{1}{e^{\beta'_R \omega'_m}- 1}H(m)+\frac{1}{e^{\beta'_L \omega'_m}- 1}H(-m)\\
  = <n_k^R(\omega'_m)>+ <n_k^L(\omega'_m)>.
\end{split}\end{equation}
The Hawking radiation are also superposition of thermal radiation of right/left sector at different temperatures, just like the BTZ black hole case.

With those temperatures, the energy and entropy can be rewritten as,
\begin{equation}\label{21j}
  S_{R/L}=\frac{\pi}{6} V T'_{R/L},\quad E'_{R/L}=\frac{\pi}{12} V T'^{2}_{R/L}.
\end{equation}
They are just the entropy and energy for phonon gas in one dimension circle with length $V=2 \pi r_+ c$, similar to the BTZ black hole case. This fact suggest that maybe the 4-dimensional Kerr black hole is essentially $(1+1)$ object \cite{bm3,carlip2}.
\section{Impact on information loss paradox}
In the above sections, we give the microscopic states of the BTZ black hole and Kerr black hole from the boundary scalar field. The microscopic states of those black holes are represented by sequences $\{m\}$ that satisfy some constraints. Different sequences correspond to different states of the black hole, and can carry the information about the formation of the black hole. And the most possible distribution is the Bose-Einstein distribution (\ref{5b}) with zero chemical potential. The Hawking radiation is just superposition of thermal radiation of right/left sector at different temperatures.

Different from Bekenstein's suggest, we consider black holes as quantum many-body systems that composed by many atoms. The thermal radiation of right/left sectors are not exactly Boson distribution, since the energy has discrete levels and also a minimal energy (\ref{8a},\ref{25a}).
\section{Conclusion}
In this paper, we derived the Hawking radiation form the boundary scalar fields. The Hawking radiation can be considered as superposition of thermal radiation of right/left sector on the horizon at different temperatures $T'_{R/L}$. The entropy and energy of black holes have the same form as phonon gas in one dimension circle, both for the BTZ black hole and the Kerr black hole.

We also discuss the impact on the information loss paradox. The information can hidden in the microscopic states of black holes, and Hawking radiation spectrum also modified due to the discreteness of the energy level.

\acknowledgments
 This work is supported by Nanhu Scholars Program for Young Scholars of XYNU.


\end{document}